# Practical Studies for Different Methods of Lunar Occultation Timing with DSLR Cameras

A. Halavati[1,2], A. Poro[1]

[1]International Occultation Timing Association-Middle East Section; a.halavati@iota-me.com
[2]Bkaraan Observatory, Kerman, Iran

**Abstract**
There are several methods for timing occultations. Many astronomers may not have access to standard video timing tools, but many of them have access to digital single-lens reflex (DSLR) cameras. In order to increase the accuracy of timing, creative methods were investigated for the DSLR camera technique. These can be a good substitute for the less accurate visual timing method. Two methods of continuous shooting and afocal filming were examined in the experimental phase, which was calculated using maximum speed sequential photography 5 shots per second, 0.1 seconds precision and 60 frames per second shooting speed resulting in 0.0083 seconds precision timing. Two different sources of time were used for video timing: Internet clock and GPS, where GPS base results were more accurate than the Internet clock.

## Introduction

Common methods in astronomical timing surveys are generally divided into two categories: Visual timing and Video timing (Poro, 2011). Visual methods require fewer tools and are more accessible to more observers but the biggest drawback is lower timing accuracy. Conversely, video timing based procedures allow for data recording and result in a higher precision of timing to an accuracy of 0.033 seconds. However, due to the high cost of video timing tools, they are less accessible to all enthusiasts. Due to the widespread use of DSLRs in the astronomical community, and in order to use this tool instead of video timing to increase the accuracy of astronomical timing, innovative ways to take advantage of their availability were examined. As a result of our investigation, this paper proposes a DSLR camera-based approach.

The reference for our measurements of the Moon's topography is related to those made by the KAGUYA (also called Selene) spacecraft in orbit around the Moon since 2007. The amount of change is calculated from the profiles produced by KAGUYA. Therefore, the calculations are based on the occultation timing by observations compared against KAGUYA profiles which is shown by what we refer to as the Observation minus Calculated (O-C) parameter (Poro, 2019).

## Observations

A series of observations were conducted for a total of 8 nights involving lunar occultations of 9 stars and 14 timings (Table 1). The tools used to perform the timing methods were a 10 inch Schmidt-Cassegrain Meade telescope mounted on a EQ6 Sky-Watcher mount, a Nikon 5300 DSLR camera, and a clear filter.

Table 1. Observation specifications, and results for O-C

| Star Name | Date | Event Type | Method | Base Time | O-C |
|---|---|---|---|---|---|
| SAO 77578 | 2019/03/14 | Disappearance | Sequential Photography | Internet Time | 0.250 |
| SAO 98162 | 2019/03/17 | Disappearance | Sequential Photography | Internet Time | 0.280 |
| SAO 118892 | 2019/04/16 | Disappearance | Afocal Videoing | GPS | 0.080 |
| SAO 187426 | 2019/04/24 | Disappearance | Afocal Videoing | GPS | -0.160 |
| | | | | Internet Time | 0.370 |
| SAO 98446 | 2019/05/11 | Disappearance | Afocal Videoing | GPS | 0.090 |
| SAO 110264 | 2019/07/24 | Reappearance | Afocal Videoing | GPS | 0.070 |
| | | | | Internet Time | 0.240 |
| SAO 185526 | 2019/08/10 | Disappearance | Afocal Videoing | GPS | 0.044 |
| | | | | Internet Time | 0.310 |
| SAO 186848 | 2019/08/11 | Disappearance | Afocal Videoing | GPS | 0.036 |
| | | | | Internet Time | 0.180 |
| SAO 186894 | 2019/08/11 | Disappearance | Afocal Videoing | GPS | 0.052 |
| | | | | Internet Time | 0.220 |

**Timing methods and Analysis**

**A) Sequential photography method**

In this method, we connect a DSLR-camera with a T-ring to telescope tube and get data by photography in a time period from a few seconds before the estimated time of the occultation until a few seconds after occultation occurs. In this time period, we try to take as many photos as possible by automatic settings. Then, by reviewing the photos, we identify the photo where the occultation occurs. We then have the exact time of occultation in hours, minutes and seconds, and we then calculate the fraction of a second by counting the photos that were taken during the second that the occultation occurred. E.g. if we have 5 photos during that second, and occultation occurs in the third photo, then the fraction of the second is determined as 0.4. Due to the automatic time adjustment, when connecting the camera to the Internet the time information is automatically started but completely dependent on the accuracy of the internet signal. The internet time is only accurate to the nearest second and this is the reason for recording as many photos as possible during each consecutive second. In order to increase the timing accuracy, we increased the number of photos taken per second. The highest number of photos taken per second with a shutter speed setting of 1/4000 second (each photo size is about 2Mb) was 5 frames per second, and the maximum interval during which photos were taken was 20 seconds with 100 images taken with a calculated timing accuracy of 0.1 seconds.

The effective factors in increasing the number of photos were as follows:
1. Using compressed formats such as JPEG instead of RAW images.
2. Reducing the shutter speed reduction to maximize the number of images per second.
3. Using memories with higher data transfer speed increases the number of images per second.
4. Reducing storage size of each photo increases the ability to gather more data per second, but must be accomplished with care in order that the star's resolution is not impaired.

The benefits of this method include:
1. Ability to record data from dimmer objects. In this method, stars were recorded down to 9th magnitude (Figure 1).
2. The camera can be connected to the T-ring interface, which makes photography easier and provides better quality photos.
3. Ability to set the clock directly by GPS and this time is embedded in each photo`s properties.

Disadvantages of this method:

1. The maximum time that the camera is capable of continuous shooting is 20 seconds. This forces the observer to take great care in determining when to start the photo sequence so as not to miss the occultation entirely.

2. There is an unequal interval between photos which increases when more photos taken sequentially. To avoid this, it is recommended to reduce the period of taking pictures.

3. Timing is less accurate. e.g. at a rate of 5 shots per second, the timing accuracy is 0.1 seconds.

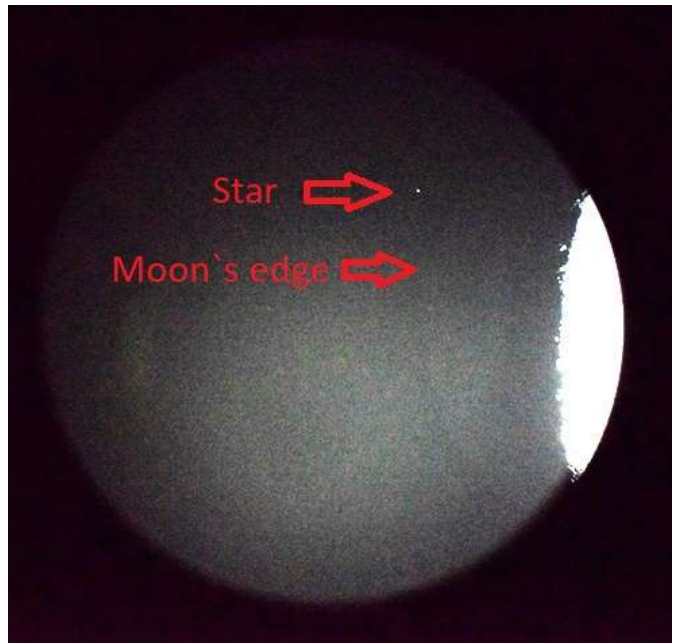

**Figure 1. The 9th magnitude star is clearly visible near the time of disappearance with the continuous shooting method**

**B) Afocal Video Method**

In this method, we use a DSLR-camera to take a video from occultation while an eyepiece attached to the telescope tube without using a T-ring. The period of data recording is from a few minutes before the estimated time of the occultation until a minute after the occultation occurs. We tooke a video from occultation throw telescope's eyepiece with afocal method. It is possible to turn on live view of camera and see the occultation instantly, and then move the camera to a basis time source such as the internet clock or a devise with GPS clock to record time on video without any interruption in videoing. It is possible to calculate occultation time accurately by frame analyze of the video. We analyzed the video using Lightwork software, at first we found the frame where the occultation occurs, Then found a frame in video that we could saw basis source of time clearly, and we call it basis time. We calculated time difference between occultation time and basis time in terms of seconds and number of frames, and reduce it form the time that is showed on the screen on basis time.

If the time source does not show a fraction of a second, the first frame in which the second is changed is used and the second fraction is set to zero.

For example, in the occultation recorded on August 10, 2019, the 6.5 magnitude star SAO185526 star was recorded and shown in Figure 2; In the left portion of this image is the last frame prior to the occultation as seen in Lightwork software; the right portion shows the moment when the occultation occurred.

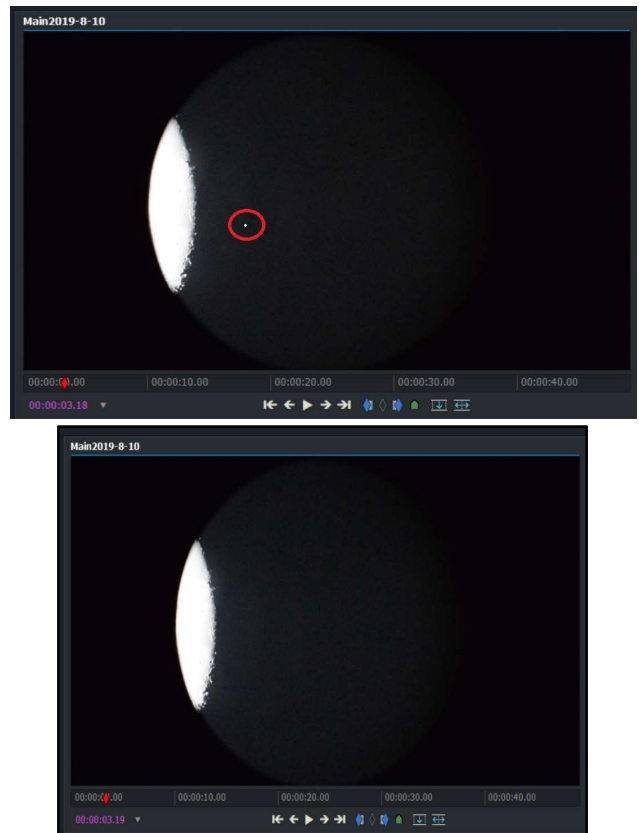

**Figure 2. A frame before the occultation occurs in Lightwork software (on the left) and the frame which occultation occurs in Lightwork software (on the right)**

Figure 3 shows the frame selected from the time sources in the video as the calculation time base. We used them to calculate occultation time with three different time sources in order to compare them and find out which one is more accurate. three different devices with different data sources were used. Two phones with SkyTiming.1 application (Lesani, 2017) and one internet connection with www.timeanddate.com were used as the most accurate time source. These were obtained using the SkyTiming.1 mobile application[1].

---

[1] http://iota-me.com/skytiming.html

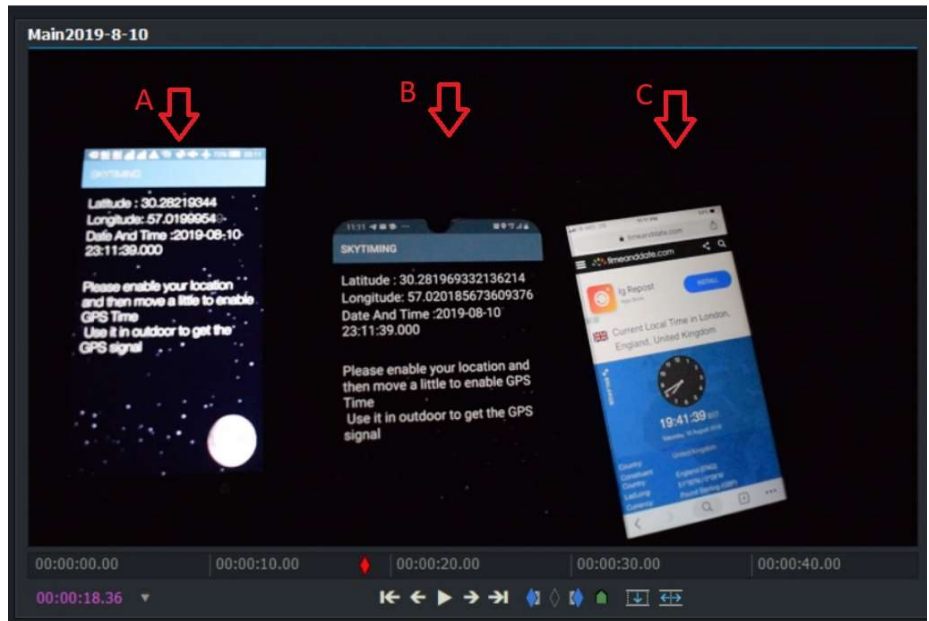

**Figure 3. Time-Based Computing Frame in Lightwork Software; A and B are showing SkyTiming app., and C is showing internet time**

The effective factors in the increased accuracy of this method are as follows:

1. Proper selection of camera parameters such as ISO, aperture according to star magnitude and Moon phase, as well as precise focus, which results in accurate recording of the time of the occultation.
2. The number of frames per second. Increasing the number of frames per second reduces the time interval between frames and ultimately increases the accuracy of computation time.
3. Time source. GPS is found as the unbitable time source, due to low errors. The time source plays an important role in increasing the accuracy of calculations. Having access to a precise timing source with, for example, filming at 60 frames per second, if the O-C time source is accurate, it should be less than 1/60 (0.16) of a second and an O-C greater than 0.16 seconds means the time source error

The benefits of this method include:

1. Ability to record data over a longer period of time. The time limit depends on the size of the camera's memory, with up to 2GB of memory available for approximately 10 minutes.
2. Higher computational accuracy due to increasing number of frames per second. In the case of shooting at 60 frames per second, it is possible to calculate the time of occurrence with an accuracy of 0.008 seconds, and in the case of shooting at 30 frames per second, calculations with an accuracy of 0.017 seconds are possible.

The disadvantages of this method include:

1. Unable to record data from dimmer objects. Due to the increase in the number of frames per second, the exposure time of each frame has been reduced and it is not possible to record data for stars greater than 8.
2. Inaccuracy of the basis time source used. In this method timing is obtained separately from the time basis source, and this data is usually within a few hundredths of a second inaccuracy. Maximum amount of error in videoing by speed of 60 frame per second should be 0.016 seconds, so higher errors is because of Inaccuracy in the basis time source used

**Conclusion**

In this project, the goal was to attempt to determine an alternative timing method to visual observation that could be proven more accurate by a rigorous testing process. Adopting such a method can be very useful for observers who do not have access to video timing tools.

The results of the 14 occultations by GPS time-lapse video capture and internet time are shown in table 1. According to our measurements, the average O-C for observations using the sequential photography method is 0.265, which is a poor result even in terms of visual timing accuracy. Generally, for observations with the afocal videoing method, the average O-C from a GPS time base is 0.0302; with an internet time base the O-C was found to be 0.2640. This shows that the accuracy of the base time has been very impressive.
In five occultations, observations did with the Afocal Videoing method, both timing (GPS and Internet time) were used, based on Table 1, which clearly shows the big difference and GPS had a reasonable result.

**References**

[1] A. Poro and P.D. Maley/2011/Astronomical Occultation book/Publisher: Daneshpajuhan Javan

[2] A. Poro, S. Memarzadeh, A. Halavati, M. Pakravanfar, I. Safaei, A. Sojoudizadeh, S. Hamedian, S. Ebadi, J. Ebrahimzadeh, A. Roshanaei4, M. Piri, M.H. Kaboli, A. Mohandes, H. Khezri, P. Eisvandi, M. Shojaatalhosseini, A. Shahdadi, M. Hesampoor, A. Gardi, K. Gholizadeh, Z. Tavangar, S. Hesampoor, S. Sarabi, M. Kazemipour, H. Hadianpour, F. Chahooshizadeh, A. Baghipour, F. Hasheminasab/2019/O-C Study of 545 Lunar Occultations from 13 Double Stars/Journal of Occultation and Eclipse (JOE), No. 6. P20-32

[3] Z. S. Lesani, A. Poro/2017/SkyTiming; A New Mobile Application for Submit and Timing/ Journal of Occultation and Eclipse (JOE), No. 4. P25-28